\documentclass[iop]{emulateapj}
\usepackage{amsmath}
\usepackage{calrsfs}

\slugcomment{Not to appear in Nonlearned J., 45.}

\shorttitle{Global Simulations of MRI}
\shortauthors{Sawai et al.}

\begin{document}

\title{Global Simulations of Magnetorotational Instability in The
  Collapsed Core of A Massive Star}

\author{H. Sawai\altaffilmark{1}, S. Yamada\altaffilmark{2},
        and H. Suzuki\altaffilmark{1}} 

\email{hsawai@rs.tus.ac.jp}

\altaffiltext{1}{Tokyo University of Science, Chiba 278-8510, Japan}
\altaffiltext{2}{Waseda University, Shinjuku, Tokyo 169-8555, Japan}

\begin{abstract}
We performed the first numerical simulations of magnetorotational
instability from a sub-magnetar-class seed magnetic field in core
collapse supernovae. As a result of axisymmetric ideal MHD
simulations, we found that the magnetic field is greatly amplified to
magnetar-class strength. In saturation phase, a substantial part of
the core is dominated by turbulence, and the magnetic field possesses
dominant large 
scale components, comparable to the size of the proto-neutron
star. A pattern of coherent chanel flows, which generally
appears during exponential growth phase in previous local simulations,
is not observed in our global simulations. While the approximate
convergence in the exponential growth rate is attained by increasing
spatial resolution, that of the saturation magnetic field
is not achieved due to still large numerical diffusion. Although the
effect of magnetic field on the dynamics is found to be mild,
a simulation with a high-enough resolution might result in a larger impact.
\end{abstract}

\keywords{supernovae: general --- magnetohydrodynamics (MHD) ---
  Instabilities --- methods: numerical  --- stars: magnetars}

\section{Introduction}
The explosion mechanism of core-collapse supernovae (CCSNe) is still
unresolved despite persistent efforts by many researchers over several
decades. Recent state-of-art simulations show that the neutrino heating
mechanism assisted by hydrodynamical instabilities revives the
accretion shock, i.e., the explosion results in. However, estimated
explosion energies are smaller than canonical 
value of order $10^{51}$~erg \citep[e.g.,][]{mar09,suw10}.

Meanwhile, effects of magnetic field on the explosion dynamics has
been studied well for the decade. Numerical simulations assuming a
strong poloidal magnetic field (typically $10^{12}$--$10^{13}$~G at
the pre-collapse phase), and rapid rotation in most cases, show that
magnetic force assists in driving the energetic explosion 
\citep[e.g.,][]{yam04,obe06,bur07,tak09,saw13}. 
The magnetic fields assumed in these simulations are so strong
that the conservation of magnetic flux during collapse results in the
field strength of 
$\gtrsim 10^{15}$~G for the proto-neutron star surface. This is
comparable to inferred surface magnetic fields of magnetar candidates
\citep[see][for review of magnetars]{woo06}.  

At present, however, the strength of the magnetic field at the pre-collapse
stage and the origin of strong magnetic field of magnetar are very
uncertain. Stellar evolution simulations by 
\citet{heg05}, which implement Tayler--Spruit dynamo model, show that 
the pre-collapse strength of the poloidal magnetic field in 15~$M_\odot$
star is only $10^{6}$~G. Meanwhile, recent observations report that
some OB stars in the main sequence stage possess an $\sim 1$~kG
surface magnetic field \citep[e.g.,][]{wad12}, which corresponds to
magnetar-class magnetic flux. \citet{fer06}
carried out a population synthesis calculation from main sequence
stars to neutron stars, assuming the magnetic flux is conserved in the
post-main-sequence evolution (fossil field hypothesis). Their result
implies that $\sim 10$\% of OB stars have magnetar-class magnetic flux,
while the majority have $\sim$1--2 orders of magnitude weaker one.

Even when the pre-collapse magnetic flux is weak, corresponding to the
magnetic field of $\lesssim 10^{13}$~G for the proto-neutron star
surface, MHD instabilities may
amplify it to magnetar-class strength. So far, there have been a
small number of works focusing on this issue. \citet{tho93} argued that
convective dynamo in proto-neutron stars generates a magnetar-class
magnetic field. Simulations performed by \citet{end12} shows that
standing accretion shock instability amplifies the magnetic field
around the proto-neutron star surface from $\sim 10^{12}$~G to
$\sim 10^{14}$~G. For rapidly rotating progenitors,
magnetorotational instability \citep[MRI,][]{bal91} may greatly
amplify the magnetic field \citep{aki03}. Local box simulations 
characterizing a post-bounce core show that an initially weak magnetic
field, $\sim 10^{12}$--$10^{13}$~G inside the proto-neutron star,
exponentially grows due to MRI  \citep{obe09,mas12}.\footnote{In the
  simulations performed by 
  \cite{obe09}, although the size of the simulation box is small,
  $\sim 1$~km, compared to the iron core, the global gradients of
  physical quantities are taken into account, by which they refer
  these simulations semi-global.}

In simulations of MRI from a sub-magnetar-class seed magnetic 
field, $\lesssim 10^{13}$~G, a quite fine grid size compared to the scale of 
the iron core, e.g., $\sim 1000$~km, is required: The wavelength of
maximum growing mode around the surface of the proto-neutron star is 
\begin{eqnarray}
\lambda_{\textrm{MGM}}&\sim&\frac{2\pi v_{\textrm{A}}}{\Omega}\nonumber\\
&\sim& 20 \textrm{m} 
\left(\frac{\rho}{10^{12}\textrm{g cm}^{-1}}\right)^{-\frac{1}{2}} 
\left(\frac{B}{10^{12}\textrm{G}}\right)
\left(\frac{\Omega}{10^{3}\textrm{rad s}^{-1}}\right)^{-1},
\end{eqnarray} where $v_{\textrm{A}}$ is Alfv\'en velocity.
In order to attain such a high resolution, local simulations are
adopted in previous works \citep{obe09,mas12}. However, in local
simulations, it is problematic to prepare a suitable background state:
Although an initially stationary background is usually used in local
simulations, post-bounce cores are dynamical. Additionally, local
simulations are incapable of studying the global dynamics. To relieve
these problems, global simulations are desirable.  

In this letter, we report the first global simulations of MRI in CCSNe
from a seed magnetic field of sub-magnetar-class flux. To ease high
computational cost demanded in global simulations, we carried out
computations in axisymmetry and in limited radial range, 50--500~km.
Particular attention is paid to, (i) whether MRI amplify the magnetic
field to magnetar-class strength even in the dynamical background
of CCSNe, and (ii) whether the amplified magnetic field is strong
enough to affect the dynamics. 

\section{Numerical Methods}
In the simulations, we solve the following ideal
MHD equations, employing a time-explicit
Eulerian MHD code, \textit{Yamazakura} \citep{saw13}:
{\allowdisplaybreaks
\begin{eqnarray}
&&\frac{\partial\rho}{\partial
  t}+\nabla\cdot(\rho\mbox{\boldmath$v$})=0\label{eq.mhd.mass},
\\ 
&&\frac{\partial}{\partial t} (\rho\mbox{\boldmath$v$})+
\nabla\cdot\left(\rho\mbox{\boldmath$v$}\mbox{\boldmath$v$}-
\frac{\mbox{\boldmath$B$}\mbox{\boldmath$B$}}{4\pi}\right)\nonumber\\
&&\hspace{1pc}=-\nabla\left(p+\frac{B^2}{8\pi}\right)-\rho\nabla\Phi 
\label{eq.mhd.mom},
\\  
&&\frac{\partial}{\partial t}\left(e+\frac{\rho
    v^2}{2}+\frac{B^2}{8\pi}\right)
\nonumber\\
&&\hspace{1pc}+\nabla\cdot\left[\left(e+p+\frac{\rho
      v^2}{2}+\frac{B^2}{4\pi}\right) 
\mbox{\boldmath$v$}\right.
\nonumber\\
&&\hspace{4pc}
\left.-\frac{(\mbox{\boldmath$v$}\cdot\mbox{\boldmath$B$}) 
\mbox{\boldmath$B$}}{4\pi}\right]
=-\rho(\nabla\Phi)\cdot\mbox{\boldmath$v$}\label{eq.mhd.eng},
\\
&&\frac{\partial\mbox{\boldmath $B$}}{\partial t}=
\nabla\times\left(\mbox{\boldmath$v$}\times\mbox{\boldmath$B$}\right)
\label{eq.mhd.far},
\end{eqnarray}}in which notations of the physical variables follow
custom. Here, $\Phi$ is Newtonian mono-pole gravitational
potential. Computations are done with polar coordinates in two
dimension, assuming axisymmetry and equatorial symmetry. We
adopt a tabulated nuclear equation of state produced by 
\citet{she98a, she98b}. Neutrinos are not dealt
with in our simulations. The electron fraction, which is required to
obtain the pressure of a fluid element from the EOS table, is given
by a prescription suggested by \citet{lie05}. We should note that the
Liebend{\"o}rfer's prescription is only valid until bounce.
\citet{saw13} discussed that adopting this prescription in the
post-bounce phase may lead to underestimation of pressure by upto 20\%.

Before simulating MRI, we first
follow the collapse of a 15~$M_\odot$ progenitor star (S. E. Woosley
1995, private communication) for the central region of 4000~km radius, 
until $\sim 100$~ms after bounce (basic run). After the central
density reaches $10^{12}$~g~cm$^{-3}$, the core is covered with
$N_r\times N_\theta = 720\times 30$ numerical grids, where the spatial
resolution is 0.4--23~km. The core is assumed to be rapidly rotating
with the initial angular velocity profile of 
\begin{eqnarray}
\Omega(r)=\Omega_{0}\frac{r_0^2}{r_0^2+r^2},
\end{eqnarray}
where $r_0=1000$~km and $\Omega_{0}=3.9$~rad s$^{-1}$, corresponding
to a millisecond proto-neutron star after collapse. The same 
dipole-like magnetic field configuration as one employed by
\citet{saw13} is initially assumed with the typical field strength of $\sim
10^{11}$~G around the pole inside a radius of 1000~km, with which the
strength of $\sim 10^{13}$~G is obtained for the proto-neutron star
surface. The initial rotational energy and magnetic energy divided by
the gravitational binding energy are $5.0\times 10^{-3}$ and
$5.3\times 10^{-7}$, respectively. 

To conduct high resolution global simulations to capture the
growth of MRI in the core, we restrict the numerical domain to a
radial range of $50<(r/$km$)<500$ (MRI run). The initial condition of
MRI run is set by mapping the data of basic run onto the above domain,
at 5~ms after bounce. At that time, the density at the inner-most grids is
$(1.5$--$2.2)\times 10^{11}$g~cm$^{-3}$. The inner and outer radial boundary  
conditions for MRI run are given by the data of basic run, except
for $B_r$. The boundary values of $B_r$ are given so that the
divergence free condition of the magnetic field is satisfied. The
grid spacing is such that the radial and angular grid sizes are same, 
viz. $\Delta r=r\Delta\theta$, at the inner and outer most
cells. We perform MRI
runs with four different grid resolutions, where the grid sizes of the
inner most cells, $\Delta r_{\textrm{min}}$, (and the numbers of grids,
$N_r\times N_\theta$,) are 12.5~m ($8900\times 6400$), 25~m
($4700\times 3200$), 50~m ($2300\times 1600$), and 100~m ($1200\times
800$). Note that with above parameters, typical MRI growth wavelength
around the inner boundary is several 100~m at the beginning of MRI
runs. We stop each MRI run, before the shock surface reaches the outer
boundary.  

\section{Results}\label{result}
\begin{figure}
\epsscale{1.2}
\plotone{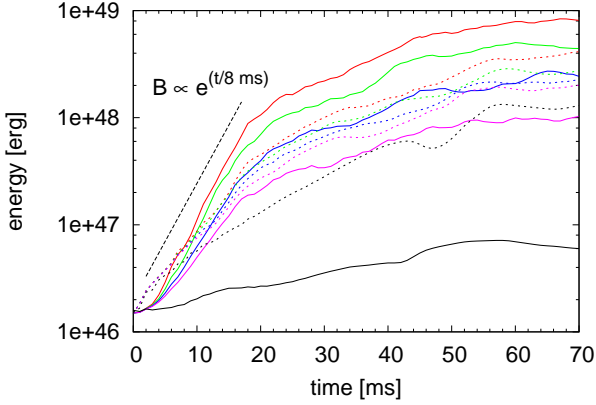}
\caption{Evolutions of the poloidal magnetic energies (solid lines)
  and the toroidal magnetic energies (dotted lines) for MRI run with
  $\Delta r_{\textrm{min}}=12.5$~m (red lines), $\Delta
  r_{\textrm{min}}=25$~m (green lines), $\Delta
  r_{\textrm{min}}=50$~m (blue lines), $\Delta
  r_{\textrm{min}}=100$~m (magenta lines), and basic run (black lines).}
\label{t-b}
\end{figure}

In each MRI run, we found a clear exponential growth of the poloidal
magnetic energy during $t\approx 4$--18~ms, where $t=0$~ms corresponds
to the beginning of MRI run (see left panel of
Figure~\ref{t-b}). The growth timescale is larger for higher grid
resolution, but almost converges to
$\tau_{\textrm{MRI,num}}\approx 8$~ms at MRI run with $\Delta
r_{\textrm{min}}=25$~m. Afterward, the  
poloidal magnetic energy gradually increases, and roughly saturates around
$t\approx 50$~ms, in which the saturation energy is larger for higher
grid resolution. The evolution of toroidal magnetic energy does not
show clear exponential growth in each MRI run. Although the growth
rate is initially larger compared with that of the poloidal magnetic
energy, it becomes smaller after the exponential growth of the
poloidal magnetic energy sets in. At saturation, in MRI run with
$\Delta r_{\textrm{min}}=12.5$~m, the poloidal magnetic energy is
twice as large as the toroidal magnetic energy. 

\begin{figure*}
\epsscale{1}
\plottwo{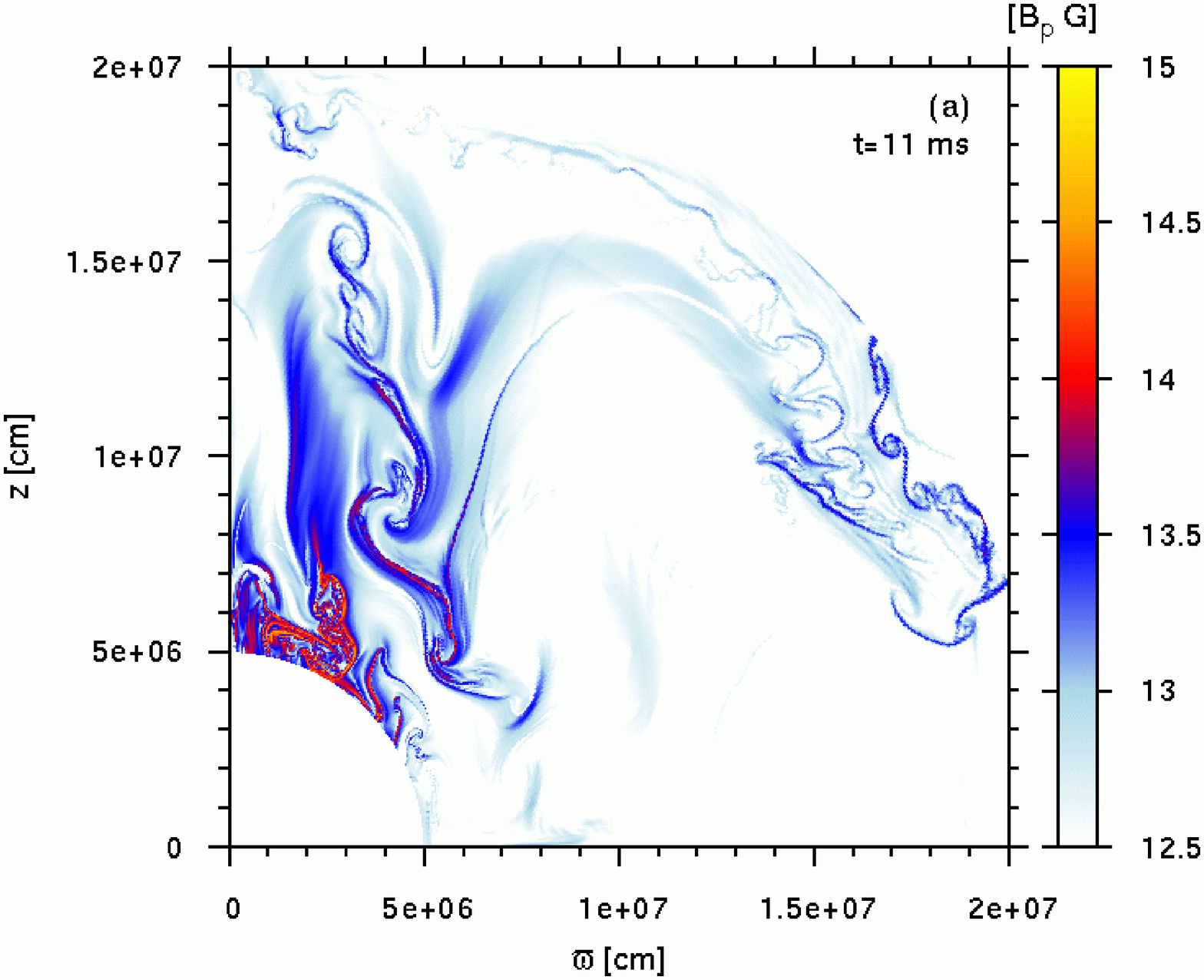}{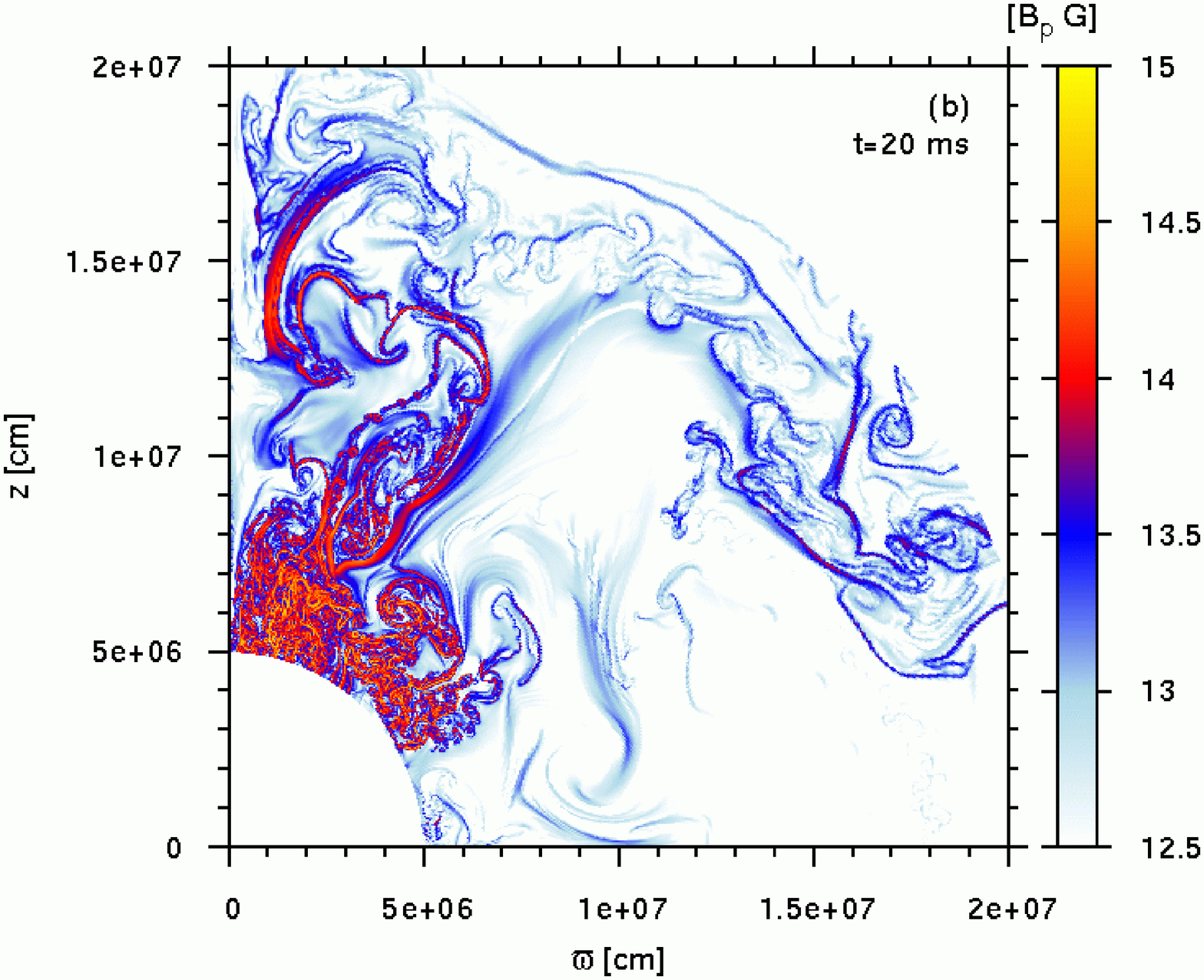}
\plottwo{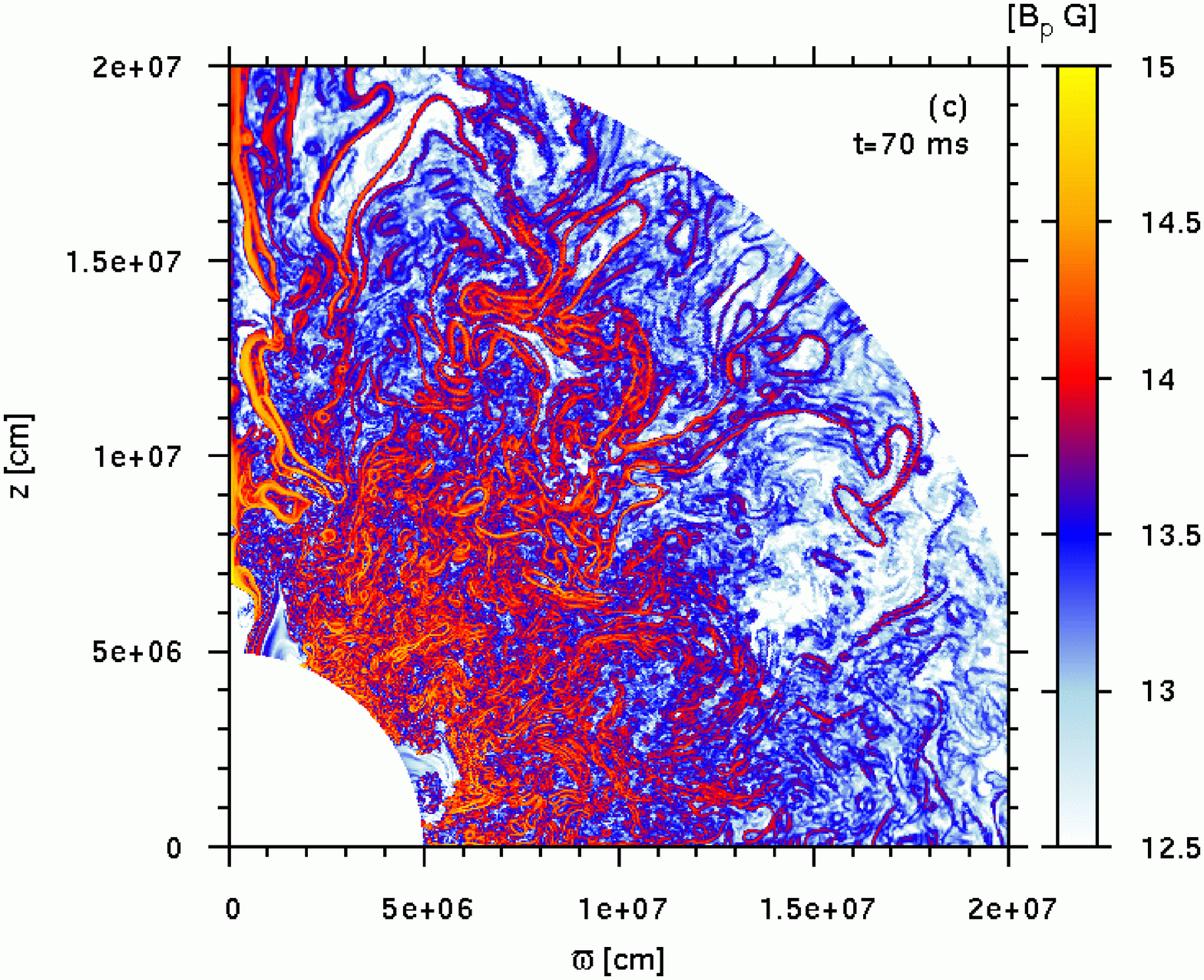}{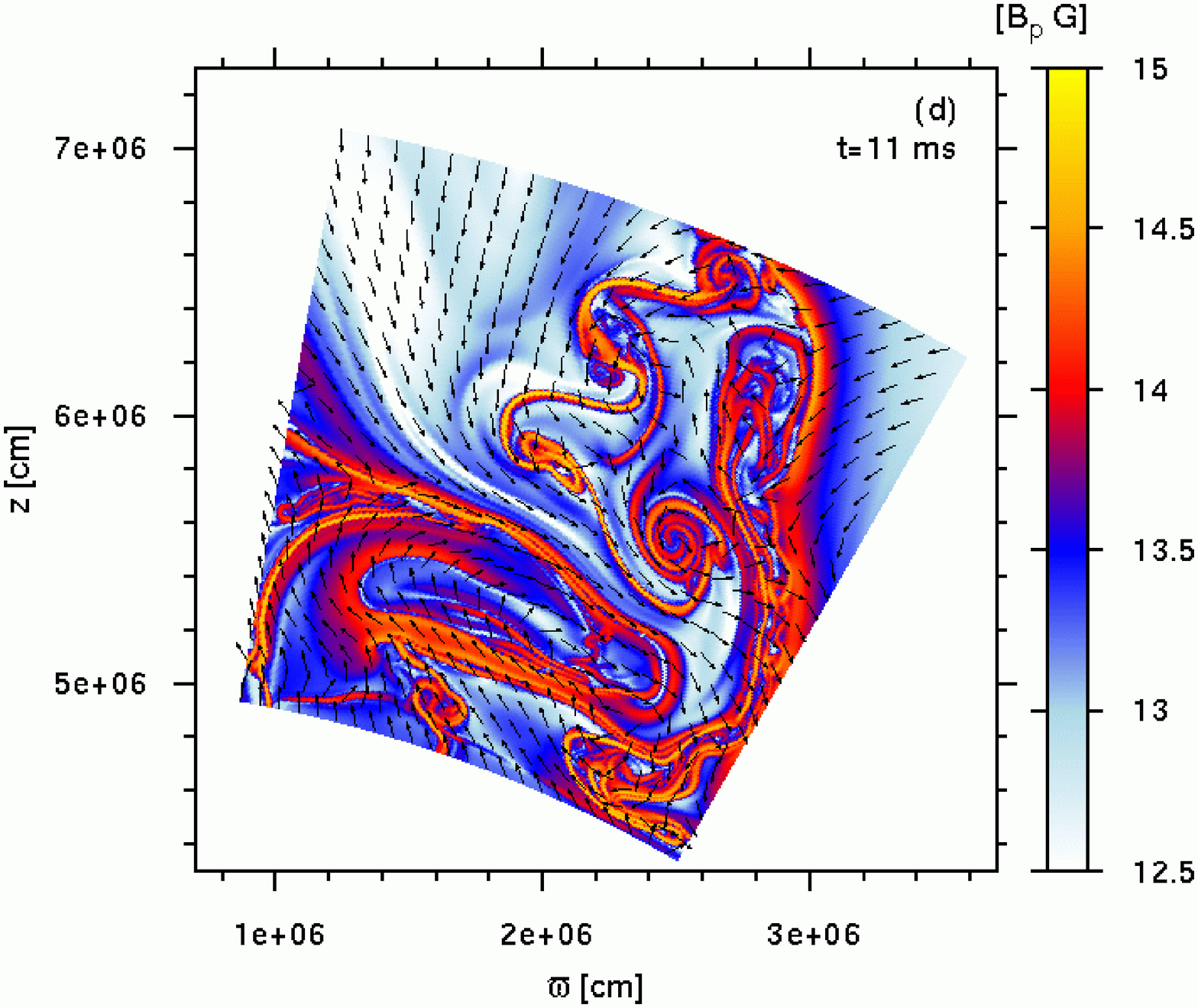}
\plottwo{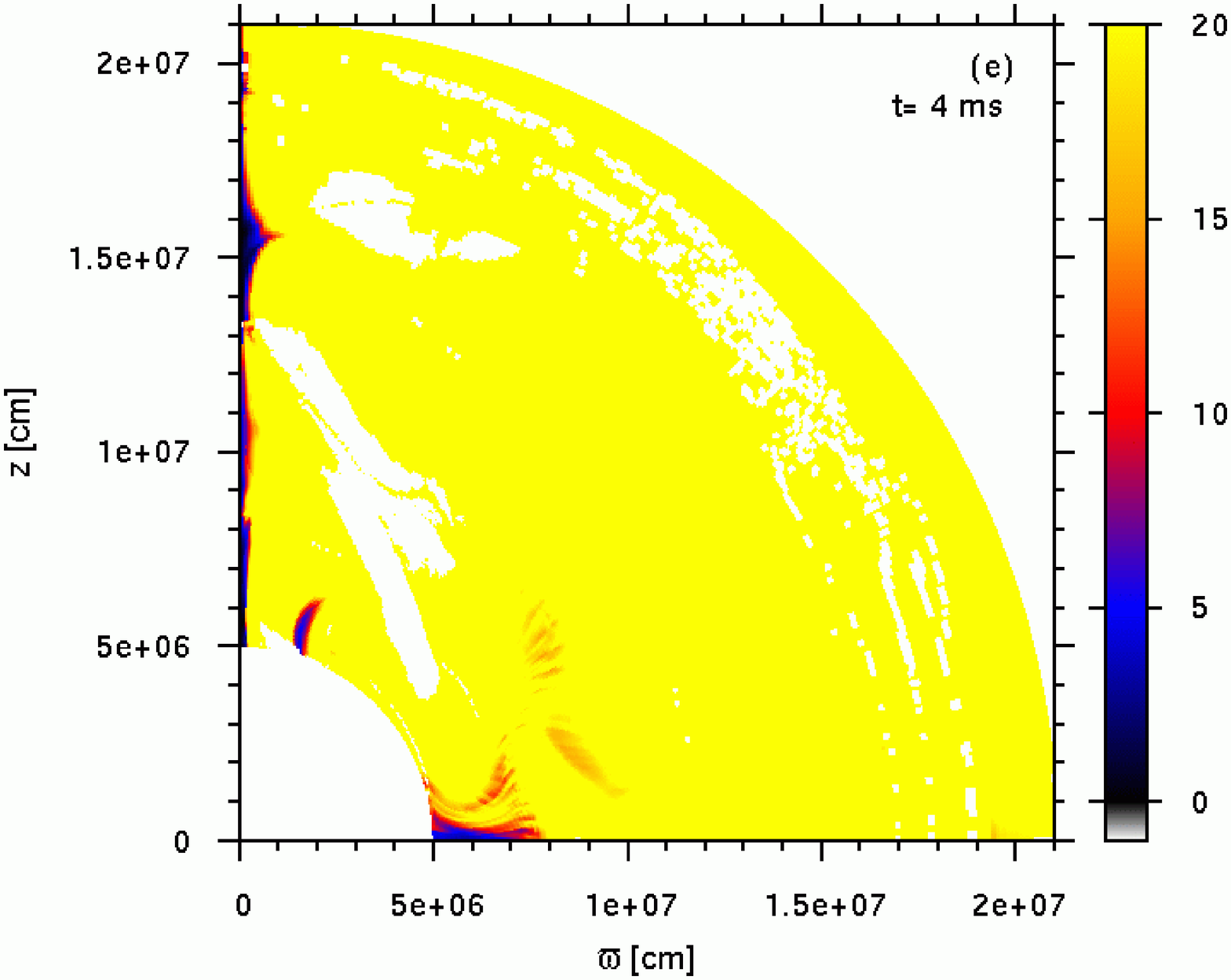}{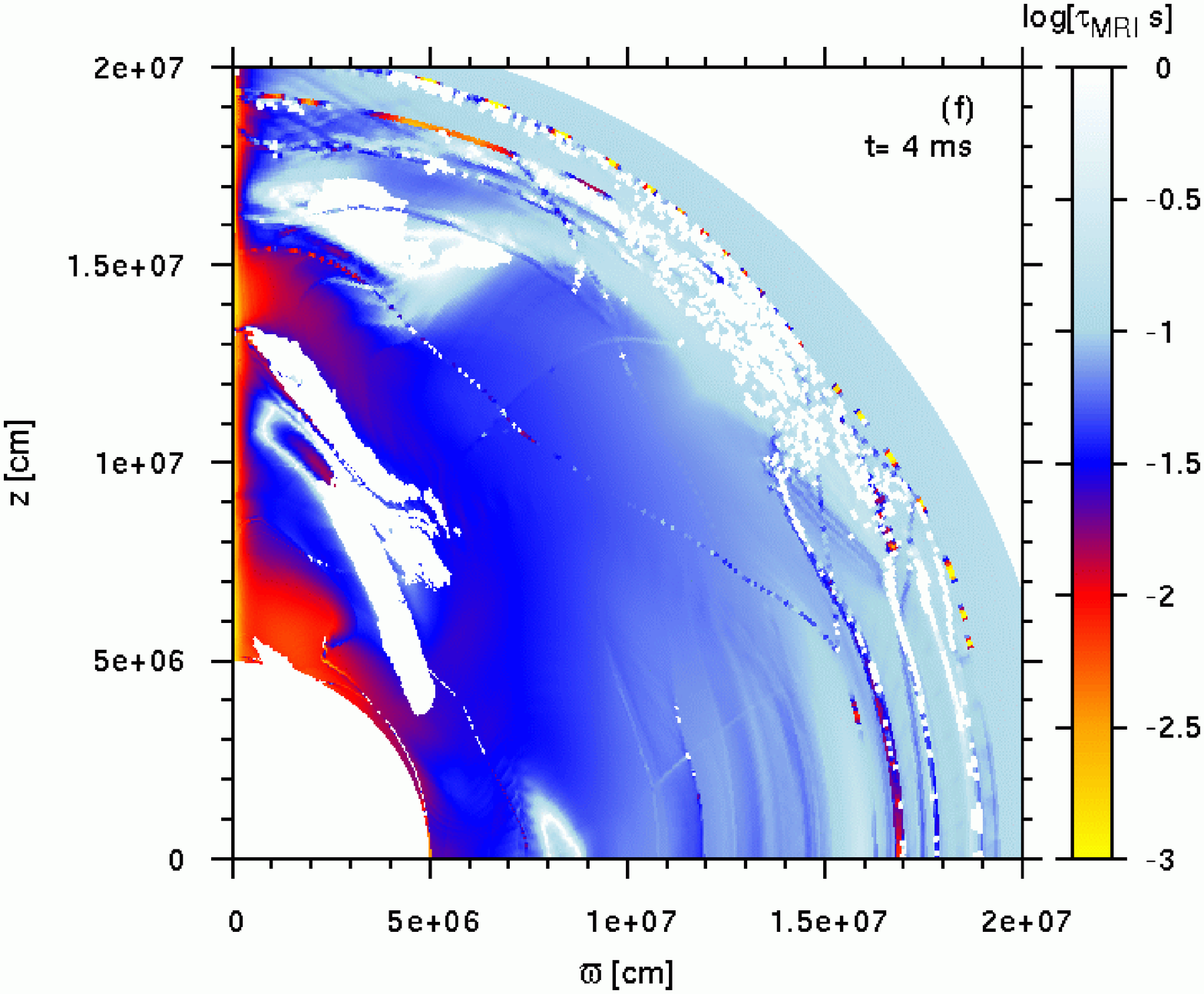}
\caption{(a)--(c): Distribution of poloidal magnetic field strength at (a)
  $t=11$~ms, (b) $t=20$~ms, and (c) $t=70$~ms. Only a part of the
  numerical domain is depicted in each panel. (d): Distribution of poloidal
  magnetic field strength (color) together with velocity directions
  (arrows) at 11~ms for $50\lesssim (r/km)\lesssim 70$ and
  $10^\circ\lesssim\theta \lesssim 30^\circ$. (e): Distribution
  of the maximum growing wavelength of MRI divided by the grid scale
  at $t=$4~ms. (f): Distribution of the maximum growing timescale of
  MRI (Equation~\ref{tmri}) at $t=$4~ms. In panel (e) and (f),
  white-colored areas include the location stable to MRI. All panels
  are depicted for MRI run with $\Delta r_{\textrm{min}}=12.5$~m.}
\label{multi2d}
\end{figure*}

We also follow the variation of the poloidal magnetic
field structure, for MRI run with $\Delta r_{\textrm{min}}=12.5$~m. 
Around $t\approx 4$~ms, when the exponential growth begins, magnetic
field lines around the pole in the vicinity of the inner
boundary start to bend. During the exponential growth phase, the
magnetic field lines in this region are further stretched, and
filaments of strong magnetic field, $\sim 10^{14}$~G, appear (see
panel (a) of Figure~\ref{multi2d} for $t=11$~ms). After the exponential 
growth ceases, the topology of magnetic field lines becomes rather
tangled, which implies that the flow becomes turbulent (see panel (b)
for $t=20$~ms). Subsequently, filaments of strong magnetic field and
the turbulent flow pattern, which first appear around the pole, also
emerge in other regions. This phase corresponds to the duration of
the gradual increase of the poloidal magnetic energy,
$t\approx 18$--50~ms, observed in Figure~\ref{t-b}. After the poloidal
magnetic energy saturates 
($t\approx 50$~ms), a considerable fraction inside a radius of 150~km is
dominated by turbulent flow, where the strength of the magnetic field 
reaches $\sim 10^{14}$--$10^{15}$~G in filamentous flux tubes (see
panel (c) for $t=70$~ms). 

The bent magnetic field lines, the exponential growth of the poloidal
magnetic field, the fact the growth rate is larger for higher gird
resolution, and the turbulence after the exponential growth, all
invoke the occurrence of MRI in our simulations. As seen in panel (e)
of Figure~\ref{multi2d}, the number of grids covering the maximum growing
wavelength in MRI run with $\Delta r_{\textrm{min}}=12.5$~m is more
than 20 in most areas, although it is less than five
in some limited locations in the vicinity of the pole and
equator. Thus, our highest-resolution MRI run seems almost capable to
fairly capture the linear growth of MRI. According to
\citet{aki03}, the growth timescale obtained by the linear theory is 
\begin{eqnarray}
\tau_{\textrm{MRI,th}}&=&2\pi\left|(\eta^2-2\eta+1)\Omega^2 + \frac{\eta-1}{2}
\left(\xi N^2+\eta\frac{d\Omega^2}{d\ln r}\right)\right.\nonumber\\
&&\left.+\frac{1}{16\Omega^2}\left(\xi N^2 + \eta\frac{d\Omega^2}{d\ln
    r}\right)^2
\right|^{-1/2}\label{tmri},
\end{eqnarray} where 
\begin{eqnarray}
\xi^2&=&(1-\sin 2\theta)^2,\\
\eta^2&=&\sin^2\theta(1-\sin 2\theta).
\end{eqnarray} 
In panel (f) of Figure~\ref{multi2d}, the distribution of
$\tau_{\textrm{MRI,th}}$ for MRI run with $\Delta
r_{\textrm{min}}=12.5$~m at 4~ms is depicted. It is found that, except
for the very vicinity of the pole, where the grid resolution is not
enough, the
growth timescale is shortest, $\tau_{\textrm{MRI,th}}\sim$10~ms, around
the pole in the vicinity of the inner boundary. This is consistent
with the fact that magnetic field amplification first occurs there.
Additionally, the numerically estimated growth timescale,
$\tau_{\textrm{MRI,num}}\approx 8$~ms 
(Figure~\ref{t-b}), approximately coincides with the above theoretical 
value. We also estimated $\tau_{\textrm{MRI,th}}$ without the effect
of the gradient of entropy and the lepton fraction, by setting $N=0$
in Equation~(\ref{tmri}), but found insignificant difference from
the original one, which indicates that convection does not much
contribute to amplify the magnetic field. With all the above facts,
we conclude that amplification of poloidal magnetic field found in our
simulations is mainly due to MRI. 

Since axisymmetry is assumed in our simulations, the toroidal
component of the magnetic field is not directly amplified by MRI. The
growth of the toroidal magnetic energy, which is slower than that of
the poloidal magnetic energy, seems due to winding of magnetic
field lines by differential rotation.

In local simulations performed by \citet{obe09}, the pattern of
coherent channel flows appears during the exponential growth
phase. In our simulations, on the other hand, such a pattern is never
observed. Although chanel-like configurations are 
locally observed in the exponential growth phase (see panel (d) of
Figure~\ref{multi2d}), they are incoherent, and soon disrupted in 
timescale of milliseconds. Such 
immediate disruption of channel flows seems to be caused by
dynamical flow motions in the background, which is difficult to be taken
into account in local simulations.

\begin{figure*}
\epsscale{1}
\plottwo{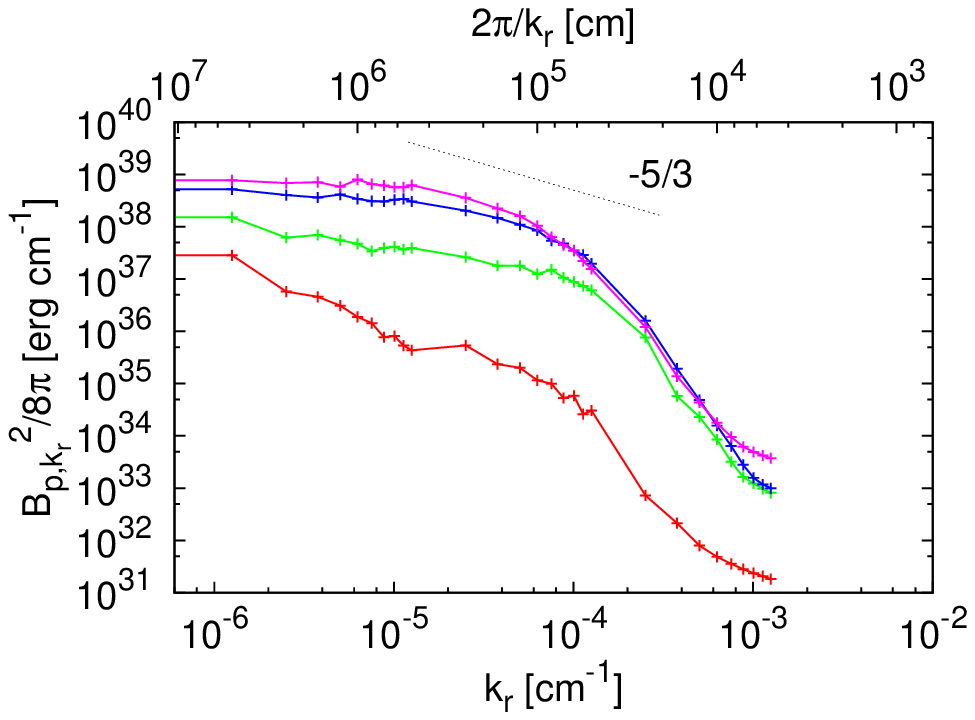}{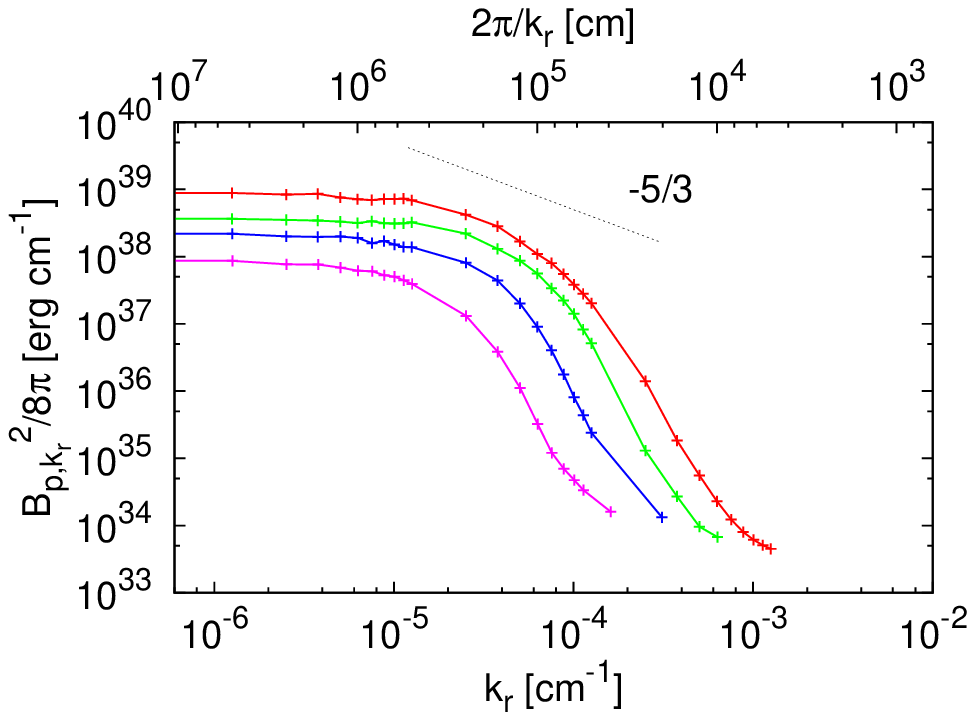}
\caption{Left: Spectra of the poloidal magnetic energy for MRI
run with $\Delta r_{\textrm{min}}=12.5$~m at $t=4$~ms (red line),
$t=11$~ms (green line), $t=20$~ms (blue line), and $t=70$~ms (magenta
line). Right: Average spectra of the poloidal magnetic energy
over $t=65$--70~ms, for MRI run with $\Delta r_{\textrm{min}}=12.5$~m
(red line), $\Delta r_{\textrm{min}}=25$~m (green line), $\Delta
r_{\textrm{min}}=50$~m (blue line), and $\Delta r_{\textrm{min}}=100$~m
(magenta line). Each spectrum is derived for radial wave numbers,
$k_r$, and for a region $50<(r/\textrm{km})<100$ and
$15^\circ<\theta<60^\circ$.} 
\label{spect}
\end{figure*}

In left panel of Figure~\ref{spect}, the energy spectra of the
poloidal magnetic energy are plotted for MRI
run with $\Delta r_{\textrm{min}}=12.5$~m at different evolutionary
phases. Each spectrum is derived for radial wave numbers, $k_r$, and
for a region $50<(r/\textrm{km})<100$ and $15^\circ<\theta<60^\circ$.
At $t=4$~ms, the spectrum shows the dominance of a large scale,
$\gtrsim 50$~km, just reflecting the structure of the background magnetic
field. During the linear growth phase, $t\approx 4$--18~ms,
smaller scale structures, $\lesssim 10$~km, grow fast compared 
with the larger ones. Consistently, we found that the maximum
growing wavelength of MRI around the pole is generally $\sim 0.1$--10~km
during this phase. At the end of the simulation (saturation phase), the
spectrum shows almost flat distribution for $r\gtrsim 5$~km, and
steep decay for $r\lesssim$1~km. In the range between them, a slope
close to $\propto k^{-5/3}$ is observed, which seems to
correspond to the inertial region of Kologomorov's theory. Such
spectrum affirms that the flow is turbulent. Also,
it indicates that large scale components of the magnetic field,
comparable to the size of the proto-neutron star, are produced
in our simulations. Since the maximum growing wavelength of MRI is
generally smaller than this scale, as mentioned above, the large scale
components are likely to be a result of inverse cascade.

We also compare the average spectra of the poloidal magnetic energy for
$t=65$--70~ms among MRI runs (right panel of Fig~\ref{spect}). It is
found that in a lower resolution run, the spectrum tends to turn the
steep decay at larger scale, which seems to be due to a larger
numerical diffusion. It is likely that this causes smaller spectral
energy for a low resolution run over whole scales, and smaller
saturation level observed in Figure~\ref{t-b}.

\begin{figure*}
\epsscale{0.5}
\plotone{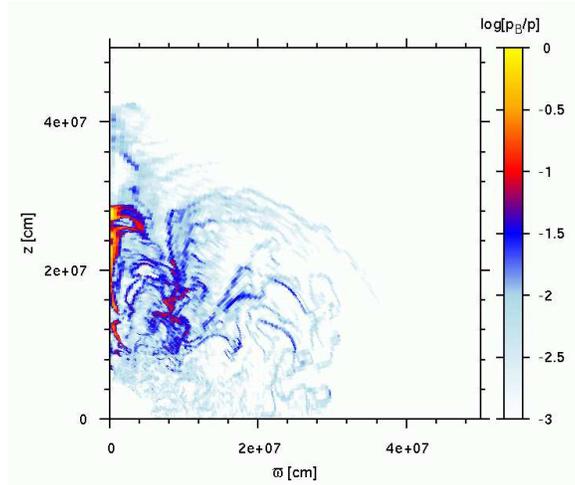}
\caption{Distribution of the ratio of magnetic pressure to matter
  pressure for MRI run with $\Delta r_{\textrm{min}}=12.5$~m at $t=70$~ms.} 
\label{ibeta}
\end{figure*}

Figure~\ref{ibeta} shows that magnetic pressure reaches $\gtrsim 10$\%
of matter pressure in some locations, which implies that the magnetic
field mildly affects the dynamics. Note, however, that we did not obtain the
convergence of the saturation magnetic field, and a larger saturation
level is attained for a higher resolution run (Figure~\ref{t-b}). This
suggest that the simulation with a high-enough resolution might result
in a larger impact of the magnetic field on the dynamics.

\section{Discussion and Conclusion}
We have carried out the first global simulations of MRI in CCSNe from a
sub-magnetar-class seed magnetic field, $\sim 10^{13}$~G around the
surface of the proto-neutron star, and rapid rotation, in which
two-dimensional ideal MHD equations are solved with assumptions of
axisymmetry and equatorial symmetry. As a result of computations, we
found that MRI greatly 
amplifies the magnetic field to magnetar-class strength, $\sim
10^{14}$--$10^{15}$~G, and that the magnetic field at saturation
phase is dominated by large scale structure. Although, the impact of
the magnetic field on the dynamics found to be mild, we do not obtain
the convergence of the saturation magnetic field strength, which is
larger for a higher resolution run. A simulation with a higher
resolution is necessary to assess the actual impact. We also found
that the evolution of flow pattern in our global simulations are quite
different from those appear in local simulations by \citet{obe09}.

Since MRI and turbulence are intrinsically
non-axisymmetric phenomena, three-dimensional simulations are,
in fact, necessary to lead to conclusive results. Note, however, that
even the world's best computers may only capable to simulate global
MRI from a magnetar-class seed magnetic field in three dimension.
\citet{obe09} found that the main difference between
two-dimensional and three dimensional simulations is that
the former results in a larger saturation magnetic field compared with
the latter. They argued that the smaller saturation level in three
dimension is caused by the disruption of coherent chanel flows before
they grow well due to
non-axisymmetric parasitic instabilities. Such difference might not
appear in global simulations, since our simulations show that coherent
chanel flows do not develop even in two dimension.

While the present simulations set the position of the inner
boundary at $r=50$~km, we found in basic run that most rotational
energy is reserved inside a radius of 50~km. It is expected that a
simulation with the inner boundary located at a smaller radius
would result in a larger saturation magnetic field. What we
obtained in this study may be a lower limit.

Although, in our simulations, a sub-magnetar-class magnetic field is
amplified to magnetar-class strength due to MRI, we should be
cautious in concluding that MRI can be the origin of magnetar fields.
Before that, further investigations may be required on the sustainability
of a large scale, strong magnetic field until, e.g., the formation of
the neutron star. Additionally, it is worth
investigating whether magnetar-class strength is
also attained by MRI from a weaker seed magnetic field, e.g.,
$<10^{12}$~G around the proto-neutron star surface.
Note that previous local simulations of MRI,
including ones in the context of accretion disks, show that a weaker
seed magnetic field results in a lower saturation level
\citep[e.g.,][]{obe09,oku11}. The dependence of the saturation magnetic field on
the rotational velocity, which also has been discussed in many local
simulations \cite[e.g.,][]{mas12}, is another issue to be studied in
future works. 

\acknowledgments
H.S. is grateful to Kenta Kiuchi for useful advise on MPI
parallelization. 
Numerical computations in this work were carried out at the 
Yukawa Institute Computer Facility.
This work is supported by a Grant-in-Aid for Scientific Research from
the Ministry of Education, Culture, Sports, Science and Technology,
Japan (20105004, 21840050, 24244036.)

\end{document}